\documentclass[aps,prl,twocolumn,preprintnumbers,showpacs,floatfix,nofootinbib]{revtex4}
\usepackage{amsmath,graphicx}

\newcommand{\mean}[1]{\langle #1 \rangle} 

\begin{document}

\preprint{BI-TP 2009/04}

\title{Medium-induced broadening and softening of a parton shower}

\author{Nicolas Borghini}
\email{borghini@physik.uni-bielefeld.de}
\affiliation{Fakult\"at f\"ur Physik, Universit\"at Bielefeld, Postfach 100131, 
  D-33501 Bielefeld, Germany}
\date{\today}

\begin{abstract}
The modifications of the angular and transverse momentum distributions of quarks
and gluons inside a parton shower due to the presence of a medium are studied 
within an analytical description that reduces to the modified leading 
logarithmic approximation (MLLA) of QCD in the absence of medium.
\end{abstract}


\maketitle

The combined measurements of hadrons and photons at large transverse momenta 
performed at the relativistic heavy-ion collider (RHIC)~\cite{d'Enterria:2009am}
are most naturally interpreted in terms of the energy degradation of high-$p_T$ 
partons in the medium formed in Au--Au or Cu--Cu collisions prior to their 
hadronization.
A significant amount of the fast parton energy is radiated away as soft gluons~\cite{Baier:2000mf,Gyulassy:2003mc,Kovner:2003zj}: the net effect of the 
medium is an enhancement of this radiation over the ``vacuum'' parton shower 
due to the Dokshitzer--Gribov--Lipatov--Altarelli--Parisi (DGLAP) parton 
evolution that leads to jets in e$^+$e$^-$ or pp/p$\bar{\rm p}$ collisions.
It is thus expected that the characteristics of a ``jet'' created in an 
ultrarelativistic heavy-ion collision\footnote{I shall herafter assume that some
  notion of a jet---whose specification might be driven by the attempts at 
  identifying collimated clusters of energy above a fluctuating 
  background in data~\cite{Putschke:2008wn,Salur:2008hs}---makes sense even in 
  the high-multiplicity environment of a nucleus--nucleus collision at high 
  energy.}
differ from those of a jet found in collisions of elementary particles.

Possible medium-induced modifications of jets include a broadening of the 
transverse momentum spectrum~\cite{Baier:1996sk,Salgado:2003rv} and the 
distortion of the longitudinal distribution inside the jet~\cite{%
  Borghini:2005em}, increasing jet multiplicities~\cite{Salgado:2003rv,%
  Borghini:2005em,Dremin:2006da,Armesto:2008qe,Ramos:2008qb}, or changes in the 
jet hadrochemistry~\cite{Sapeta:2007ad}.
Assessing these effects is an experimental challenge~\cite{Putschke:2008wn,%
  Salur:2008hs,Alessandro:2006yt,D'Enterria:2007xr,Grau:2008ef}, which would 
give access to properties of the medium probed by the energetic parton.

The satisfactory description of data on the longitudinal and transverse momentum
distributions of hadrons inside a ``vacuum'' jet are successes of perturbative 
QCD, specifically of its modified leading logarithmic approximation (MLLA), 
which takes into account the leading double logarithmic terms together with 
single logarithmic contributions~\cite{Dokshitzer:1988bq,Khoze:1996dn}, and of 
next-to-MLLA corrections that keep track of energy conservation of parton 
splittings with increased accuracy~\cite{Arleo:2007wn}.
Therefore, it would be desirable to be able to describe in-medium jets with a 
formalism that in the absence of a medium reduces to (N)MLLA.
In particular, the probabilistic interpretation of the latter in terms 
independent successive branchings with angular ordering due to destructive 
interferences inside the parton shower makes it highly suited to a Monte-Carlo 
implementation. 
Since energetic partons created in a heavy-ion collision propagate in an 
expanding medium with an anisotropic geometry, a quantitative characterization 
of the resulting partonic shower will require a numerical approach:
the interest of a ``medium-modified'' MLLA is here again obvious.

While the first numerical implementations of in-medium parton showers are 
appearing~\cite{Zapp:2008gi,Armesto:2008qh}, it is still necessary to have 
analytical calculations---obviously under simplified assumptions---of possible 
properties of the modified showers, as references against which Monte-Carlo 
computations can be tested.
In this Letter, I shall thus present results on the angular and transverse 
momentum distributions of partons inside a medium-modified shower.
Following the approach proposed in Ref.~\cite{Borghini:2005em}, the influence 
of the medium is modeled as an enhancement of the singular $1/z$ parts of the 
leading-order parton splitting functions by a constant factor $1+f_{\rm med}$ 
(this factor is denoted $N_s$ in Refs.~\cite{Dremin:2006da,Ramos:2008qb}):
\begin{eqnarray}
 & \displaystyle\label{modified_Pgq}
  P_{gq}(z)=C_F\bigg[\frac{2(1+f_{\rm med})}{z}-2+z\bigg], & \\
 & \displaystyle\label{modified_Pgg}
  P_{gg}(z)=2C_A\bigg[\frac{1}{1-z}+\frac{1+f_{\rm med}}{z}-2+z(1-z)\bigg], &
\end{eqnarray}
while the other two splitting functions remain unchanged.
The modeling is admittedly crude, yet has the advantage of allowing controlled 
analytical calculations, while mimicking the increased radiation of soft gluons 
of the approaches~\cite{Baier:2000mf,Gyulassy:2003mc,Kovner:2003zj} that 
describe RHIC results on leading hadron suppression.
Additionally it incorporates the conservation of energy at each step of the 
parton shower evolution, which plays possibly an important role in shaping the 
global jet characteristics.
A more realistic change of the splitting functions, which accounts for the 
coherent Landau--Pomeranchuk--Migdal radiation in a thermalized weakly coupled 
deconfined medium can be found in Ref.~\cite{Arnold:2008zu}. 
Reference~\cite{Armesto:2007dt} provides a somewhat equivalent description in 
terms of Sudakov form factors.

Consider a jet with energy $E$ and opening angle $\Theta_0$.
For any jet particle with 4-momentum $(k_0,\vec{k})$ one defines
\begin{equation}
\label{defs_l&y}
\ell\equiv\ln\frac{E}{k_0}=\ln\frac{1}{x}, \quad
y\equiv\ln\frac{k_\perp}{Q_0}\approx\ln\frac{k_0\sin\Theta}{Q_0},
\end{equation}
where $Q_0$ is the infrared cutoff parameter and $k_\perp\geq Q_0$ resp.\ 
$\Theta\leq\Theta_0$ is the transverse momentum resp.\ angle of the radiated 
parton with respect to the energy flow direction (``jet axis'').
Finally, let $Y_{\Theta_0}\!\equiv\ln(E\sin\Theta_0/Q_0)$, corresponding to the
maximum transverse momentum of a particle in the jet.
While approximating $\sin\Theta\sim\Theta$ is often made for the small 
opening angles relevant for jets in high-energy collisions, in the following I 
shall retain the sine.
Additionally, I shall take $Q_0=\Lambda_{\rm QCD}\simeq250$~MeV, which gives a 
good agreement with data for longitudinal hadron spectra, and show results for 
the values $Y_{\Theta_0}\!=5.1$ and $Y_{\Theta_0}\!=6$, corresponding to jets with 
energies typical for RHIC (40~GeV) and LHC (100~GeV), respectively.

The starting point to study the angular distribution inside a jet is the 
distribution $F_{\!A_0}^h\!(x,\Theta,E,\Theta_0)$ of the energy fraction $x$ of 
hadrons within a subjet with an opening angle $\Theta<\Theta_0$~\cite{%
  Dokshitzer:1988bq}.
$F_{\!A_0}^h$ is the convolution of the probability 
$D_{\!A_0}^A(u,E\Theta_0,uE\Theta)$ that the initial parton $A_0$ emit parton 
$A$ with the energy fraction $u$ and the virtuality $uE\Theta$ and of the 
fragmentation function $D_{\!A}^h(x/u,uE\Theta,Q_0)$ for the production of 
hadron $h$ off $A$ with the energy fraction $x/u$ and a transverse momentum 
scale $uE\Theta\geq Q_0$:
\begin{eqnarray}
\lefteqn{F_{\!A_0}^h(x,\Theta,E,\Theta_0)\equiv} & \cr 
 & \displaystyle\sum_{A=q,g}
  \int_x^1\!{\rm d}u\,D_{\!A_0}^A\!(u,E\Theta_0,uE\Theta)\,
   D_{\!A}^h\Big(\frac{x}{u},uE\Theta,Q_0\Big).\label{F_A0^h}
\end{eqnarray}
The number of hadrons $h$ inside the opening angle $\Theta$ is given by the 
integral of this distribution over $x$~\cite{Ochs:2008vg}:
\begin{eqnarray}
\lefteqn{\hat{N}_{\!A_0}^h(\Theta,E,\Theta_0)\equiv} & \cr 
 & \displaystyle\sum_{A=q,g}
  \int_{Q_0/E\Theta}^1\!{\rm d}u\,uD_{\!A_0}^A\!(u,E\Theta_0,uE\Theta)\,
   N_{\!A}^h(uE\Theta,Q_0),\quad\label{hatN_A0^h}
\end{eqnarray}
where $N_{\!A}^h(uE\Theta,Q_0)$ denotes the multiplicity of hadrons inside the 
``subjet'' initiated by parton $A$ with the virtuality $uE\Theta$.
The physical picture of Eqs.~(\ref{F_A0^h}-\ref{hatN_A0^h}) is that, as a 
consequence of the angular ordering of the splitting processes, the hadrons that
end up in the subjet originate from quarks and gluons emitted in the shower when
its typical $k_\perp$ scale was $uE\Theta$, after an evolution from the initial 
scale down to that intermediate scale.

Since the purpose in this Letter is to establish the modifications of jets 
induced by a medium compared to their ``vacuum'' MLLA properties, for the 
hadronic fragmentation function $D_{\!A}^h$ relevant for vacuum jets I shall use 
the corresponding MLLA (partonic) function $D_{\!A}^{\rm MLLA}$, in the limit 
$Q_0\to\Lambda_{\rm QCD}$.
The latter is in the case of a gluon jet related to the MLLA ``limiting 
spectrum''~\cite{Dokshitzer:1988bq} through 
$\tilde{D}^{\lim}(\ell,y)=xD_g^{\rm MLLA}(\ell,y)$.
Note that exact energy conservation is spoiled in the approximations leading to 
$D_{\!A}^{\rm MLLA}$ and thus $\tilde{D}^{\lim}$;
yet the latter still represents a good baseline at sufficiently small $x$.
Consequently, the hadron multiplicity inside a gluon jet is given by
\begin{equation}
\label{g-jet_multiplicity}
N_g^h(E\Theta,Q_0)=
  {\cal K}^h Y_\Theta^{-B/2+1/4}\exp\sqrt{\frac{16N_c}{\beta_0}\,Y_\Theta}\,,
\end{equation}
and that inside a quark jet by~\cite{Mueller:1983cq}
\begin{equation}
\label{q-jet_multiplicity}
N_q^h(Y_\Theta) = \frac{C_F}{N_c}
  \left(1+\frac{a-3N_c}{4N_c}\sqrt{\frac{4N_c}{\beta_0Y_\Theta}}\right)
  N_g^h(Y_\Theta)\,,
\end{equation}
where $Y_\Theta\equiv\ln(E\sin\Theta/Q_0)$, $a=(11N_c+2N_f/N_c^2)/3$, 
$\beta_0=(11N_c-2N_f)/3$ and $B=a/\beta_0$, while ${\cal K}^h$ is, following the
local parton-hadron duality (LPHD) hypothesis, a proportionality constant 
relating partonic and hadronic properties.
The actual value of ${\cal K}^h$ will be unimportant in the following, since I
shall assume that the same value should hold for vacuum and in-medium jets. 

The longitudinal $\ln(1/x)$ distribution of partons inside a (gluon) jet within 
the adopted modeling of medium effects was reported in Ref.~\cite{%
  Borghini:2005em}, and amounts to a strong distortion of the MLLA hump-backed 
plateau.
This leads to increased (sub)jet multiplicities, $\beta_0$ in Eqs.~(\ref{%
  g-jet_multiplicity}-\ref{q-jet_multiplicity}) being replaced by 
$\beta_0/(1+f_{\rm med})$ and $a$ by $a+4N_cf_{\rm med}$.

\begin{figure}[t]
\includegraphics*[width=\linewidth]{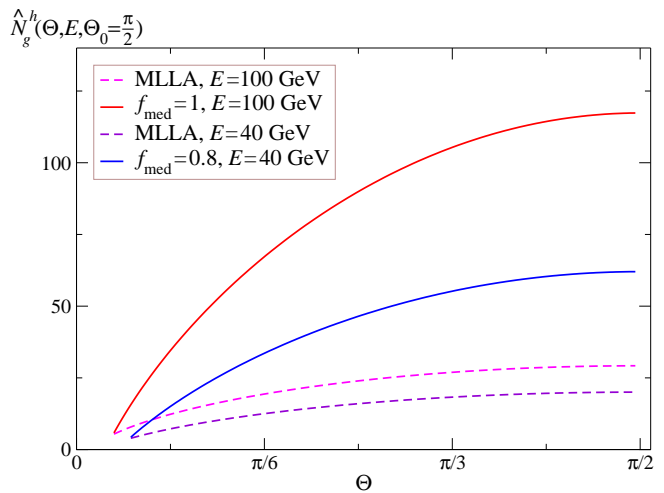}
\vspace{-4mm}
\caption{\label{fig:ThetaDist}Angular dependence of multiplicities inside gluon 
  jets with $E=100$~GeV ($Y_{\Theta_0}=6$) in MLLA and for $f_{\rm med}=1$, and 
  with $E\simeq 40$~GeV ($Y_{\Theta_0}=5.1$) in MLLA and for $f_{\rm med}=0.8$.}
\vspace{-4mm}
\end{figure}

Regarding $D_{\!A_0}^A$, for vacuum jets it is given by the usual expression 
following from the Altarelli--Parisi parton splitting functions. 
Replacing the latter with the modified functions~(\ref{modified_Pgq}-\ref{%
  modified_Pgg}) and performing a Mellin transformation as well as an expansion
of $N_{\!A}^h$ around $u=1$ in Eq.~(\ref{hatN_A0^h}), one obtains readily the 
multiplicity $\hat{N}_{\!A_0}^h$,\footnote{The detailed calculations will be 
  reported elsewhere.} which is plotted vs.\ the opening 
angle $\Theta$ in Fig.~\ref{fig:ThetaDist} for both MLLA and medium-distorted 
jets.
The value $f_{\rm med}=0.8$, which allows to reproduce a nuclear modification 
factor $R_{AA}\approx 0.2$~\cite{Borghini:2005em}, was adopted for the RHIC 
energy, while $f_{\rm med}=1$ was chosen for the higher (LHC) energy.
As in Ref.~\cite{Ochs:2008vg}, the LPHD constant was taken to be 
${\cal K}^h=0.2$.
Figure~\ref{fig:ThetaDist} shows that the gradient 
${\rm d}\hat{N}_g^h/{\rm d\Theta}$ is larger over a broader angular range for 
the medium-modified than for the MLLA gluon jet. 
This behavior, which is also observed on quarks jets (not shown here), is the 
in-medium {\em angular broadening\/} of the parton shower which was observed
in Q-{\sc Pythia}~\cite{Armesto:2008qe}.
Note that in Fig.~\ref{fig:ThetaDist} the low-$\Theta$ region has been 
suppressed, since even the MLLA results are not reliable for small angles (see 
the remark on low $k_\perp$ values below).

The hadron multiplicities $\hat{N}_g^h$ represented in Fig.~\ref{fig:ThetaDist} 
include {\em all\/} hadrons down to the infrared cutoff $Q_0$. 
Given the high-multiplicity environment of heavy-ion collisions, experimental 
studies will most probably rather investigate the modifications of jet 
properties above some $p_T$ (with respect to the beam axis) cutoff which might 
be quite large.
In that case, the angular broadening is significantly less marked.%
  \footnotemark[\value{footnote}]
\smallskip

Let me turn to the distribution of transverse momenta $k_\perp$ in a jet.
Deriving Eq.~(\ref{F_A0^h}) with respect to $\ln\Theta$---or equivalently, at 
fixed $x$, $\ln k_\perp$---yields the double differential single-particle 
inclusive distribution of hadrons inside a jet with opening angle $\Theta_0$ and
energy $E$~\cite{Dokshitzer:1988bq}:
\begin{equation}
\label{2-diff_dist}
\frac{{\rm d}^2N}{{\rm d}x\,{\rm d}\ln k_\perp} \simeq
\frac{{\rm d}^2N}{{\rm d}x\,{\rm d}\ln\Theta} =
\frac{\rm d}{{\rm d}\ln\Theta} F_{\!A_0}^h\!(x,\Theta,E,\Theta_0).
\end{equation}

Within a small-$x$ approximation, one can expand $D_{\!A}^h$ around $u=1$ in 
Eq.~(\ref{F_A0^h}). 
This allows one, after some algebra~\cite{PerezRamos:2005nh}, to rewrite 
$F_{\!A_0}^h$ in terms of the average color current $\mean{C}_{\!A_0}$ inside a 
jet initiated by parton $A_0$ and of the distribution in $\ell=\ln(1/x)$ of 
hadrons inside a gluon jet, 
$\bar{\cal D}_g(\ell,y)\equiv xD_g^h(x,E\Theta,Q_0)$.
Inserting the result in Eq.~(\ref{2-diff_dist}) yields
\begin{equation}
\label{2-diff_dist-bis}
\bigg(\frac{{\rm d}^2N}{{\rm d}\ell\,{\rm d}y}\bigg)_{\!\!g,q}\! =
\frac{\rm d}{{\rm d}y}
  \bigg[\frac{\mean{C}_{g,q}}{N_c}\bar{\cal D}_g(\ell,y)\bigg].
\end{equation}
This expression was computed for vacuum jets within MLLA~\cite{%
  PerezRamos:2005nh}, using the limiting spectrum $\tilde{D}^{\lim}$ for 
$\bar{\cal D}_g$, and later including some next-to-MLLA corrections, which 
yield improved agreement with measurements by the CDF Collaboration for 
$k_\perp>1$~GeV/$c$~\cite{Arleo:2007wn}.
Given the exploratory approach pursued in this Letter, I shall only investigate 
the influence of the medium modifications in Eqs.~(\ref{modified_Pgq}-\ref{%
  modified_Pgg}) on the MLLA $k_\perp$-distributions.
The average color currents $\mean{C}_{g,q}$ following from the modified 
splitting functions can be computed\footnotemark[\value{footnote}] and inserted
in Eq.~(\ref{2-diff_dist-bis}) together with the distorted 
longitudinal spectrum~\cite{Borghini:2005em}.
For values of $\ell$ where the assumptions behind the MLLA computation are 
fulfilled (see discussion in Ref.~\cite{PerezRamos:2005nh}), the resulting 
double differential single-particle distribution is, with respect to the vacuum 
one, depleted at low $k_\perp$ ($y\lesssim 1.5$) and significantly larger
at higher $k_\perp$~\cite{Borghini:2009fj}.
The growth at high $k_\perp$ obviously reflects the angular broadening already 
observed in Fig.~\ref{fig:ThetaDist}, while the low-$k_\perp$ behavior should 
probably be considered with care, since it corresponds to a region where MLLA is
not to be trusted. 

\begin{figure}[t]
\includegraphics*[width=\linewidth]{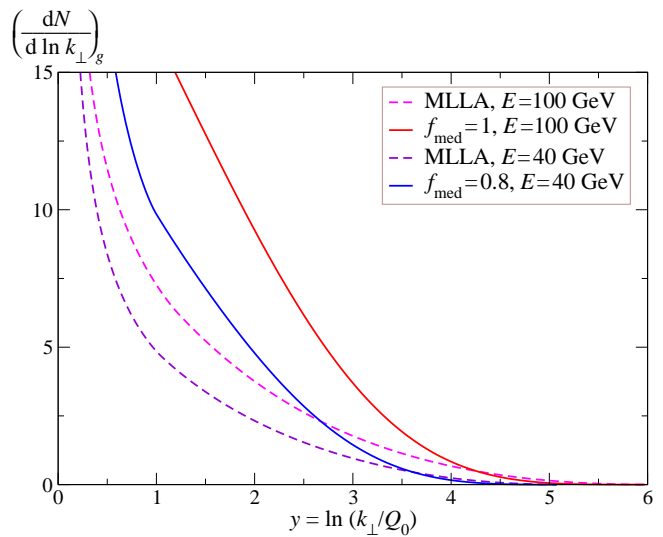}
\vspace{-4mm}
\caption{\label{fig:lnkTdistributions}$\ln k_\perp$-distribution inside gluon 
  jets for the same cases as in Fig.~\ref{fig:ThetaDist}.}
\vspace{-4mm}
\end{figure}

Integrating Eq.~(\ref{2-diff_dist-bis}) over $\ell$ yields the transverse 
momentum distribution inside a jet:
\begin{equation}
\bigg(\frac{{\rm d}N}{{\rm d}\ln k_\perp}\bigg)_{\!\!g,q}\! =
  \int\!{\rm d}\ell\,
  \bigg(\frac{{\rm d}^2N}{{\rm d}\ell\,{\rm d}\ln k_\perp}\bigg)_{\!\!g,q}.
\vspace{-1mm}
\end{equation}
The latter is displayed in Fig.~\ref{fig:lnkTdistributions} for vacuum and 
medium-distorted gluon jets at RHIC and LHC energies.
While the values below $y=\ln(k_\perp/Q_0)=1.5$, should not be taken too 
literally---the divergence at $y\to 0$, which reflects that of the running QCD
coupling constant $\alpha_s(k_\perp)$ when $k_\perp\to\Lambda_{\rm QCD}$, hints 
at the breakdown of the perturbative regime---, the overall trend of medium 
effects seems to be a push of the distribution towards larger transverse 
momenta:
this is a {\em transverse momentum broadening\/} of the jets. 
Note that the ``usual'' source of transverse momentum broadening (which was 
however not seen in the first {\sc Jewel} simulations~\cite{Zapp:2008fy}) is 
the transfer of momentum to the jet in the interactions between the latter and 
the medium~\cite{Baier:1996sk}: this is not taken into account in the model. 
The phenomenon reported here is different, since no momentum is injected into 
the jet from the outside. 
The eventual broadening of the jet will result from both effects, which do not 
necessarily add up constructively. 
Investigating that issue is beyond the scope of the present study

\begin{figure}[t]
\includegraphics*[width=\linewidth]{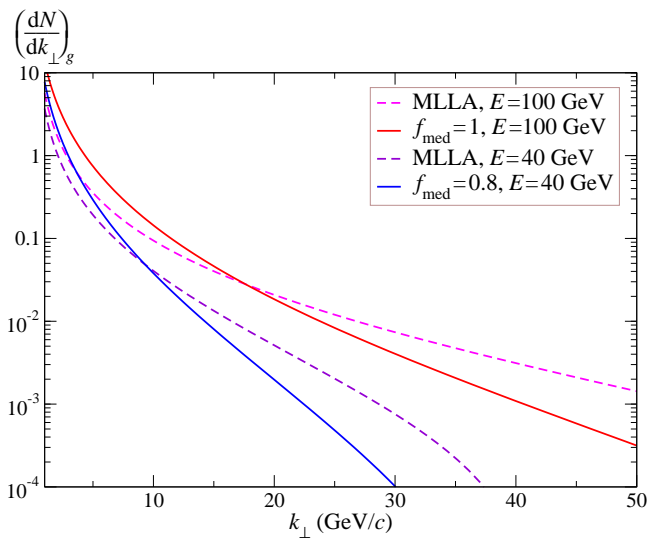}
\vspace{-4mm}
\caption{\label{fig:kTdistributions}Transverse momentum distribution inside
  gluon jets for the same cases as in Fig.~\ref{fig:ThetaDist}.}
\vspace{-2mm}
\end{figure}

Since the effect of the medium is to redistribute partons from high towards
low $x$-values~\cite{Borghini:2005em}, it was not totally obvious that this 
would at the same time involve a wider $k_\perp$ distribution as seen in 
Fig.~\ref{fig:lnkTdistributions}.
Both modifications can simultaneously happen only if the presence of the medium 
makes the transverse momentum spectrum steeper.
This is illustrated in Fig.~\ref{fig:kTdistributions}, which indeed shows a
medium-induced depletion of the yield at high $k_\perp$, which was barely 
visible on Fig.~\ref{fig:lnkTdistributions}, together with the enhancement at 
low $k_\perp$.
All in all, the $k_\perp$ spectrum is {\em softened\/}, which was also seen in 
Q-{\sc Pythia}~simulations~\cite{Armesto:2008qe}.

Integrating the $k_\perp$ spectrum above a given $k_\perp^{\rm cut}$ yields the 
multiplicity above that cutoff, which might be easier to assess experimentally.
One can then form some ``medium modification factor'' by dividing the 
multiplicity in a medium-distorted jet by that inside a vacuum jet. 
This ratio is shown for both gluon and quark jets in Fig.~\ref{%
  fig:RAA(k>k_cut)}:
\begin{figure}[t]
\includegraphics*[width=\linewidth]{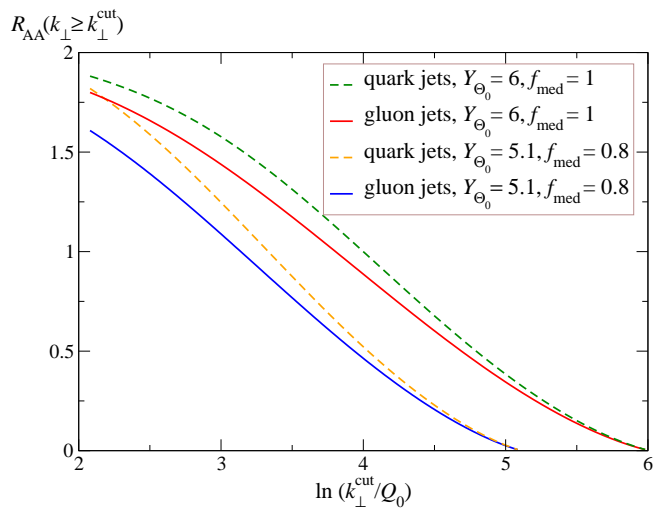}
\vspace{-4mm}
\caption{\label{fig:RAA(k>k_cut)}Ratio of the multiplicities above a given 
  $k_\perp$ cutoff inside medium-distorted and vacuum jets, for gluon and 
  quark jets at typical LHC ($Y_{\Theta_0}=6$) and RHIC ($Y_{\Theta_0}=5.1$) 
  energies.}
\vspace{-4mm}
\end{figure}
it is larger than 1 up to $k_\perp^{\rm cut}\simeq 10-15$~GeV/$c$, and is 
systematically larger for quark than for gluon jets.
\smallskip

Using an analytical formalism, I have computed the angular and transverse 
momentum distributions of partons inside a medium-distorted parton shower. 
The effect of the medium is, as anticipated, a broadening of the distributions 
with respect to those of ``vacuum'' MLLA jets, as well as a softening of the 
$k_\perp$ spectrum. 
The multiplicity over a lower $k_\perp$ cutoff, which is more easily accessible
experimentally, shows an enhancement in medium-modified jets up to large cutoff 
values.

Strictly speaking, the results obtained here hold at the parton level. 
Using the LPHD hypothesis, which for (N)MLLA jets leads to a good agreement with
measurements of $\ln(1/x)$ and $k_\perp$ spectra, they would translate into 
characteristics at the hadronic level.
The hadronization process might however possibly totally reshuffle the 
distributions inside the jet~\cite{Zapp:2008gi}, in which case the predictions 
could only be checked in Monte-Carlo simulations.
\medskip

\begin{acknowledgments}
I wish to thank Redamy P\'erez-Ramos for enlightening discussions.
\end{acknowledgments}

\end{document}